# Orbital Fulde-Ferrell-Larkin-Ovchinnikov state in an Ising superconductor

Puhua Wan[1], Oleksandr Zheliuk[1,2], Noah F. Q. Yuan[3], Xiaoli Peng[1], Le Zhang[1], Minpeng Liang[1], Uli Zeitler[2], Steffen Wiedmann[2], Nigel Hussey[2], Thomas T. M. Palstra[4], and Jianting Ye[1*]

[1] *Device Physics of Complex Materials, Zernike Institute for Advanced Materials, University of Groningen, 9747 AG Groningen, The Netherlands*

[2] *High Field Magnet Laboratory (HFML-EMFL), Radboud University, Toernooiveld 7, 6525 ED Nijmegen, Netherlands*

[3] *School of Science, Harbin Institute of Technology, Shenzhen 518055, China*

[4] *Nano Electronic Materials, University of Twente, Drienerlolaan 5, 7522 NB Enschede, The Netherlands*

**In superconductors possessing both time and inversion symmetries, the Zeeman effect of an external magnetic field can break the time-reversal symmetry, forming a conventional Fulde-Ferrell-Larkin-Ovchinnikov (FFLO) state characterized by Cooper pairings with finite momentum[1,2]. In superconductors lacking inversion symmetry, the Zeeman effect may still act as the underlying mechanism of FFLO states by interacting with spin-orbit coupling (SOC). Specifically, the interplay between the Zeeman effect and Rashba SOC can lead to the formation of more accessible Rashba FFLO states that cover broader regions in the phase diagram[3–5]. However, when the Zeeman effect is suppressed due to the spin-locking in the presence of Ising-type SOC, the conventional FFLO scenarios are no longer effective. Instead, an unconventional FFLO state is formed by coupling the orbital effect of magnetic fields with SOC, providing an alternative mechanism in superconductors with broken inversion symmetries[6–8]. Here we report the discovery of such an orbital FFLO state in multilayer Ising superconductor 2$H$-NbSe$_2$. Transport measurements show that the translational and rotational symmetries are broken in the orbital FFLO state, providing the hallmark signatures of finite momentum cooper pairings. We establish the entire orbital FFLO phase diagram, consisting of normal metal, uniform Ising superconducting phase, and a six-fold orbital FFLO state. This study highlights an alternative route to finite-momentum superconductivity and provides a universal mechanism to prepare orbital FFLO states in similar materials with broken inversion symmetries.**

# Main

In a conventional superconductor, when spatial inversion and time-reversal symmetries are respected, electrons form Cooper pairs with opposite momenta and spins as described by the Bardeen–Cooper–Schrieffer (BCS) theory (Fig. 1a). Breaking either of the two symmetries can lead to unconventional Cooper pairing with nontrivial spin or momentum configurations. The absence of spatial inversion symmetry, for example, can result in a Rashba or Ising-type spin-orbit coupling (SOC), which enhances the paramagnetic limiting field beyond the Pauli limit $B_\text{P}$ along particular directions[9–11]. Breaking time-reversal symmetry by the Zeeman effect of an external magnetic field, on the other hand, can lead to the Fulde-Ferrell-Larkin-Ovchinnikov (FFLO) state, where Cooper pairs acquire a nonzero momentum $q$[1,2] (Fig. 1b). This 'conventional' FFLO state can be stabilized in a clean superconductor[12,13] through a first-order phase transition from the BCS phase at low temperatures $T < T_\text{FFLO} \equiv 0.56 T_\text{c0}$ and high fields $B > B_\text{FFLO} \equiv 0.75 \Delta_0 / \mu_\text{m}$. Here $T_\text{c0}$ is the critical temperature at the zero $B$ field, $\Delta_0$ the pairing potential at zero field and zero temperature, and $\mu_\text{m}$ is the electron magnetic moment[14,15].

Simultaneously breaking the inversion and time-reversal symmetries are predicted to lead to an even richer manifold of finite-momentum pairing states beyond conventional FFLO[4–7,16]. In a Rashba-type superconductor, SOC polarizes the in-plane spin configuration. A much weaker Zeeman effect from a parallel external field can then deform the Fermi surface and stabilize finite-momentum pairing (Fig. 1c). In this way, Rashba SOC can extend the FFLO state to higher temperatures and lower magnetic field strengths relative to the conventional FFLO limits defined above[4,5,16].

Despite the symmetry differences, the Zeeman effect configures both the conventional and Rashba FFLOs. Considering that the orbital and Zeeman effects are simultaneously exerted by an external magnetic field, favoring the Zeeman effect requires the suppression of the orbital effect. Crucially, the Cooper pairs should not be disrupted by orbital depairing before the Zeeman effect has driven them through an FFLO transition. This explains why FFLO states have been widely reported in heavy-fermion[17,18] or low-dimensional superconductors[19–23] where the orbital effect is weakened.

In contrast to the conventional wisdom, we show here that the interlayer orbital effect from an external $B$ field can assist FFLO formation in an Ising superconductor where the Zeeman effect is suppressed. In transition metal dichalcogenides (TMDs), local symmetry breaking in individual layers induces strong Ising SOC, locking spins alternately at the $K/K'$ valley of the Brillouin zone. This strong locking effect suppresses the change of spin configuration of Cooper pairs under a parallel external field[10,11], prohibiting the Zeeman effect from driving the system into conventional FFLO states. This broken local symmetry is globally restored in a multilayer Ising superconductor with $2H$ stacking. As shown in Fig. 1d, a weak interlayer orbital effect can shift the energy bands and stabilize a finite momentum pairing with opposite momentum $q$ in adjacent layers[6]. Furthermore, this orbital FFLO state has a real-space oscillating phase in the order parameter (Fig. 1e). Compared with the conventional FFLO, the new mechanism can realize the FFLO state in field strengths substantially smaller than $B_P$, and at a temperature closer to $T_{c0}$, thereby covering a much broader proportion of phase diagram[6,7]. We noticed that an orbital-field-driven Fulde–Ferrell pairing state in a moiré bilayer TMD was studied in a recent theoretical work[24].

Here, we report the experimental realization of the orbital FFLO state in a multilayer Ising superconductor ($2H$–$NbSe_2$). The finite-momentum pairing emerges below a tri-critical point at $T^* = 0.84T_{c0}$ and $B^* = 0.36B_P$, where the orbital effect of the parallel $B$ field couples superconducting layers through the Josephson interaction. Translational and rotational symmetries are broken across the tricritical point as revealed by transitions in the vortex dynamics and the in-plane anisotropy, respectively. These broken symmetries provide the hallmark signatures of finite-momentum Cooper pairs induced by the formation of an orbital FFLO state. Following the characteristic symmetry behavior, we map out the boundary of the first-order transition and establish the full phase diagram of the states.

## Phase transition to an Orbital FFLO state

We prepare multilayer $NbSe_2$ flakes covered by $h$-BN flakes to ensure high-quality transport with a large residual resistivity ratio (*RRR*), which is on par with bulk single crystals[25]. As shown in Fig. 1g, the *RRR*, defined as $R(280\text{ K})/R(8\text{ K})$, reaches 28 for a 17 nm thick flake. A sharp superconducting transition is observed at $T_{c0} = 6.9$ K, following the charge density wave (CDW) transition at $T_{CDW} \approx 32$ K. The angular dependences of magnetoresistance are measured by orientating the external $B$ field to the 2D crystal planes. Figure 1h shows the upper critical fields, $B_{c2,\perp}$ and $B_{c2,\parallel}$, measured down to 0.3 K. Here, the $B_{c2}$ is determined as the $B$ field where $R = 0.5R_N$ (Fig. S6 of SI). From the $B_{c2,\perp}(T)$, we estimate the Pippard coherence length $\xi_0 = 11.2$ nm, which is smaller than the estimated mean free path $l_m \approx 30$ nm in our device, and comparable to that reported in high-quality bulk crystals ($l_m = 27$ nm)[26]. Satisfying $l_m > \xi_0$ locates our superconducting states within the clean regime (SI 1).

In the $B_\parallel$ configuration, $B_{c2,\parallel}(T)$ follows a square-root dependence near $T_{c0}$, consistent with a 2D GL description (Fig. 1h) for the upper critical field, $B_{c2,\parallel} = \frac{\sqrt{3}\phi_0}{\pi\xi_\parallel(0)d}(1-\frac{T}{T_{c0}})^{\frac{1}{2}}$, of a 2D superconductor with thickness $d$ (Extended data Fig. 1). With decreasing $T$, $B_{c2,\parallel}(T)$ shows a conspicuous upturn absent in either bulk[27] or few-layer thin flakes[11,28]. Similar upturns can also be observed in other high-quality flakes with intermediate thicknesses of tens of nanometers (Fig. 1i).

With the increase in NbSe$_2$ thickness from the monolayer to the bulk, the extrapolated $B_{c2,\|}(0\text{ K})$ varies gradually from $> 6B_P$ to $\sim B_P$[11,27,28]. This significant decrease of $B_{c2,\|}$ is caused by a gradual increase of interlayer Josephson coupling $J$ as the flakes get thicker: from $J = 0$ in a monolayer to its maximum in the bulk[28]. In thicker flakes, the increase of $J$ significantly weakens the Ising protection in individual layers[29]. At a thickness of 17 nm (Fig. 1h), $B_{c2,\|} = 1.5B_P$ is found to be intermediate between $B_{c2,\|}$ values reported in the bulk[27] and bilayers[28], indicating a reduced $J$ compared with the bulk crystals. It is worth noting that the upturn of $B_{c2,\|}$ is absent in bilayer NbSe$_2$[28] and our 12 nm sample with the increased disorder (Extended Data Fig. 1), confirming that high-quality transport is a crucial prerequisite for realizing the orbital FFLO state.

The upturn at $T^* = 5.8$ K ($0.84T_{c0}$) and $B^* = 4.7$ T ($0.36B_P$), as shown in Fig. 1h, is consistent with the theoretical prediction for the tricritical point of the orbital FFLO state[6,7]. Note that $T^*$ is considerably higher than the $0.56T_{c0}$ required for the conventional FFLO state[14,15]. Moreover, $B^*$ is much smaller than $B_P$ since the Zeeman effect, required for forming a conventional FFLO state, is replaced by a new mechanism based on interlayer orbital coupling when strong orbital depairing is absent in $B_\|$. Guided by $T^*$ and $B^*$, we then examine several hallmark signatures related to spatial symmetry, including vortex dynamics[30,31] and anisotropy transition[32,33], to distinguish the FFLO from the uniform phase.

## Translational symmetry breaking in the FFLO state

When the FFLO state is established by interlayer orbital coupling under $B_\|$[34], tilting $B$ fields away from the parallel direction can cause a transition from the FFLO to the uniform phase, providing a preliminary identification of the FFLO state. Figures 2a and b show distinctive polar angle $\theta$ dependencies of $B_{c2}$ across the critical point ($T^*$, $B^*$). At $T = 6.3$ K, above $T^*$, $B_{c2}(\theta)$ exhibits a single-cusp centered at $\theta = 0°$, which agrees with a 2D Tinkham fit for a uniform superconductor (Fig. 2a). At $T = 4.8$ K below $T^*$, $B_{c2}(\theta)$ shows a sharper enhancement within a critical angle $|\theta_c| \sim 1°$ (Fig. 2b). Tentatively, the dependence of $B_{c2}(\theta)$ below $\theta_c$ can be described by an additional Tinkham fit with a larger $B_{c2,\|}$. When orbital depairing is suppressed at $|\theta| < \theta_c$, the FFLO state is favorable (Fig. 2c) since it lowers the free energy compared with the uniform phase. Consequently, the FFLO transition across $T^*$ can cause an anomalous enhancement of $B_{c2}$[20,22,35].

Stabilizing the orbital FFLO state breaks the uniform order parameter of a multilayer NbSe$_2$ by adding alternating phase modulation in the out-of-plane direction (Fig. 1e), which is predicted to leave distinctive features in the vortex dynamics[30,32]. Due to the local translational symmetry breaking of the superconducting order parameter, the interlayer vortex motion is pinned in the FFLO state. To probe the vortex dynamics in the orbital FFLO state, we measure $B_{c2,\parallel}$ as a function of the Lorentz force, $\mathbf{F} = \mathbf{I} \times \mathbf{B}$, which points towards the out-of-plane direction. The associated $\mathbf{F}$ drives interlayer vortex motion in the uniform phase, reducing $B_{c2}$ via dissipation[36]. At the phase boundary between metal and orbital FFLO states, which is of second order, thermal fluctuations destroy the long-range coherence, forming a coexisting metal/orbital FFLO state. The vortex pinning can "lock" $B_{c2}$ against $\mathbf{F}$ as a fingerprint of the oscillating order parameter. As shown in the insert of Fig. 2b, the vortex pinning effect manifests itself as a change in the critical field strength $\Delta B_{c2} = B_{c2}(\varphi = 65°) - B_{c2}(\varphi = 0°)$, where $\varphi$ is an azimuthal angle within the conducting plane (Extended Data Fig. 4). $\mathbf{F}$ is weaker at $\varphi = 65°$ compared with $\varphi = 0°$, which is later shown in Fig. 3c-f. As shown in the lower panel of the insert in Fig. 2b, $\Delta B_{c2} \sim 0$ T at $\theta = 0°$ and quickly reaches $\sim 0.4$ T for $|\theta| > \theta_c$. Within a small tilt angle $\theta$, $\mathbf{F}$ can be regarded as constant. Therefore, the result $\Delta B_{c2}(\theta) = 0$ indicates a strong vortex pinning effect, surpassing the dominance of the Lorentz force due to the oscillating phases in the FFLO state. Such a pinning effect is absent in the uniform superconducting phase when $|\theta| > \theta_c$, as indicated by a finite $\Delta B_{c2}$.

Figures 2d and e show the $\Delta B_{c2}(T)$ following the normal/superconducting phase boundary, measured at $\theta = 3° > \theta_c$ and $\theta = 0° < \theta_c$, respectively. Since $B_{c2}$ increases with the decrease of temperature, at a constant $\mathbf{I}$, the $\mathbf{F}$ also increases at lower temperatures. At $\theta = 3°$, where the order parameter is uniform, $\Delta B_{c2}$ shows clear activation behavior due to the increase of $\mathbf{F}$ at lower temperatures. When $\theta = 0°$, $\Delta B_{c2}$ coincides with that measured at $\theta = 3°$ for $T > T^*$. Comparing Fig. 2d and e, the deviation in $\Delta B_{c2}$ becomes visible for $T \leq T^*$ because of the pinning effect. These contrasting behaviors confirm $T^*$ as the critical temperature to enter the orbital FFLO state.

The abrupt vortex pinning below $\theta_c$ at $T < T^*$ is consistent with the alternating phase vector $\mathbf{q}$ formed in the FFLO order parameter, switching polarities within a single unit cell (Fig. 2e). This phase configuration is analogous to the ground state of a Josephson $\pi$ junction, where the net tunneling of a supercurrent through the junction is zero. Effectively, the FFLO order parameter prohibits vortex motion driven by $\mathbf{F}$ in the out-of-plane direction[37]. As a result, upon entering the FFLO phase below $T^*$, we observe that the pinning effect increases as $T$ decreases, as shown in Fig. 2e.

## Rotational symmetry breaking in the orbital FFLO state

The orbital FFLO state also breaks rotational symmetry in the basal plane; another hallmark signature to be measured by electrical transport[31]. In this work, the anisotropy manifests itself in the azimuthal angular dependence of the magnetoresistance $R(\varphi)$. To resolve this anisotropy, which is later identified as only ~1% of the gap size, we must eliminate the effect of canting angle $\gamma$ as discussed in Methods. Figure 3c shows the azimuthal angular $\varphi$ dependence of $R_{\parallel}$ in parallel magnetic fields along the phase boundary ($T$, $B_{c2,\parallel}$) after correcting $\gamma$ (see Methods and Extended Data Fig. 3). As shown in Fig. 3d-i, the anisotropy of $R_{\parallel}$ for different phases across the $T^*$ shows a two- to six-fold transition. For 2D angular mapping above the $T^*$, $R(\varphi, \theta)$ oscillates at 180° in a two-fold anisotropy exhibiting maxima and minima for the $\boldsymbol{B} \perp \boldsymbol{I}$ and $\boldsymbol{B} \parallel \boldsymbol{I}$ configurations, respectively (Fig. 3d-f). This directional dependence in $\boldsymbol{I}$ is further characterized in Fig. 3j when $\boldsymbol{I}$ is rotated by ~90° using orthogonal pairs of electrodes. Rotating $\boldsymbol{I}$ by ~90° causes a shift of two-fold anisotropy for ~95° with a small deviation of 5° caused by electrodes (Extended Data Fig. 4). This significant shift indicates that the two-fold anisotropy has an extrinsic origin. The Lorentz force $\boldsymbol{F}$ becomes zero for $\boldsymbol{B} \parallel \boldsymbol{I}$ and reaches a maximum for $\boldsymbol{B} \perp \boldsymbol{I}$. The anisotropy of $\boldsymbol{F}$ affects the dissipative motion of vortices, causing two-fold anisotropy as shown in Fig. 3d-f and j, which is consistent with the report on bulk crystal[25] but different from another report on ultrathin $NbSe_2$[38].

As shown in Fig. 3g-i, the emergence of the six-fold anisotropy is coincide with $T^*$ (Fig. 3h), where $R(\varphi, \theta)$ reaches its minima when the $B$ field is applied along the crystalline direction of $NbSe_2$ at $\varphi = -60°, 0°$, and $60°$. In contrast to the two-fold anisotropy due to $\boldsymbol{F}$, which shifts as $\boldsymbol{I}$ is rotated, applying an orthogonal $\boldsymbol{I}$ causes no shift in the six-fold anisotropy (Fig. 3i and k), indicating its intrinsic origin. From the $R_{\parallel}(\varphi)$ variation in Fig. 3c, we quantify the anisotropy as 0.8% of the largest superconducting gap at the $K/K'$ pockets of $NbSe_2$ (SI 3). We also confirm the universal existence of the six-fold anisotropy in another comparable high-quality sample with intermediate thickness, showing an upturn in $B_{c2}$ (Extended Data Fig. 5). In contrast, similar controlled measurements carried out on ionic-gated $MoS_2$ exhibit uniform superconductivity consistent with the picture of monolayer Ising superconductivity without interlayer interaction (SI 4).

## Phase diagram of the orbital FFLO

To resolve the first-order phase boundary between the FFLO and the normal Ising superconducting phase, we measure the critical current density $J_c$ as a function of $T$ and $B_{\parallel}$ as shown in Fig. 4a. Taking the measurement at 2.5 K as an example, we find that as $J_c$ decreases with $B_{\parallel}$, a clear upturn appears at $B_{\parallel} = 5.5$ T. The kink becomes smoothed out when the temperature is increased to 6 K. In contrast, these kinks are not observed in the temperature dependence of $J_c$ when $B_{\parallel} = 0$, which rules out the multi-gap scenario as the cause of the upturn (Methods). Like the abrupt increase of $B_{c2}$ (Fig. 2b) observed at the phase boundary, entering the orbital FFLO phase at a fixed temperature from the uniform phase also lowers the free energy[32,39], enhancing $J_c$. Therefore, we determine the phase boundary as the $B$ field at which the upturn in $J_c$ occurs (Fig. 4b). The extrapolation from the measured phase boundary aligns well with the tri-critical point (the yellow dot), which is differently determined from the phase boundary by the upturn in $B_{c2}(T)$ shown in Fig. 1h.

Figure 4b shows the entire phase diagram consisting of normal metal, uniform Ising superconductor, and the six-fold orbital FFLO phase. Based on symmetry considerations for a multilayer NbSe$_2$ under $B_\parallel$, the distinctive change from isotropic to six-fold basal plane anisotropy (Fig. 3d-i) around the tri-critical points ($T^*$, $B^*$) can be described by the Ginzburg-Landau free energy shown in the Methods. We found that the critical field $B^* = \sqrt{2Jc}/b$, separating the zero momentum uniform phase and the orbital FFLO phase, is nearly temperature independent as consistently shown in the phase diagram (Fig. 4b). At high fields above the tricritical point, Cooper pairs in the $l_{\text{th}}$ layer acquire alternating finite momentum $(-1)^l \boldsymbol{q}$, where $\boldsymbol{q} = \frac{b}{2c}(\boldsymbol{B} \times \boldsymbol{z})$, to minimize the free energy. Due to the constraints of crystal rotation symmetries, the nonzero momentum $q$ couples anisotropically to the $B$ field, manifesting itself in higher-order terms (see Methods). For $B < B^*$, the $q = 0$, and the $T_c = T_0 + \frac{b^2}{4ac}(2|B^*|^2 - B^2)$ is isotropic. When $B > B^*$, where $q \neq 0$, the $T_c = T_0 + \frac{b^2}{4ac}\frac{|B^*|^4}{B^2} + \lambda(B) \cos(6\varphi)$. This six-fold anisotropy of $T_c$ is captured precisely by measuring the $R(\varphi)$ in the coexisting state (Fig. 3g-i).

Our work demonstrates the first example of an orbital FFLO phase based on an interlayer-coupled Ising superconductor. Similar orbital FFLO phases are expected following this general mechanism in many other multilayer Ising superconductors.

# Figure 1

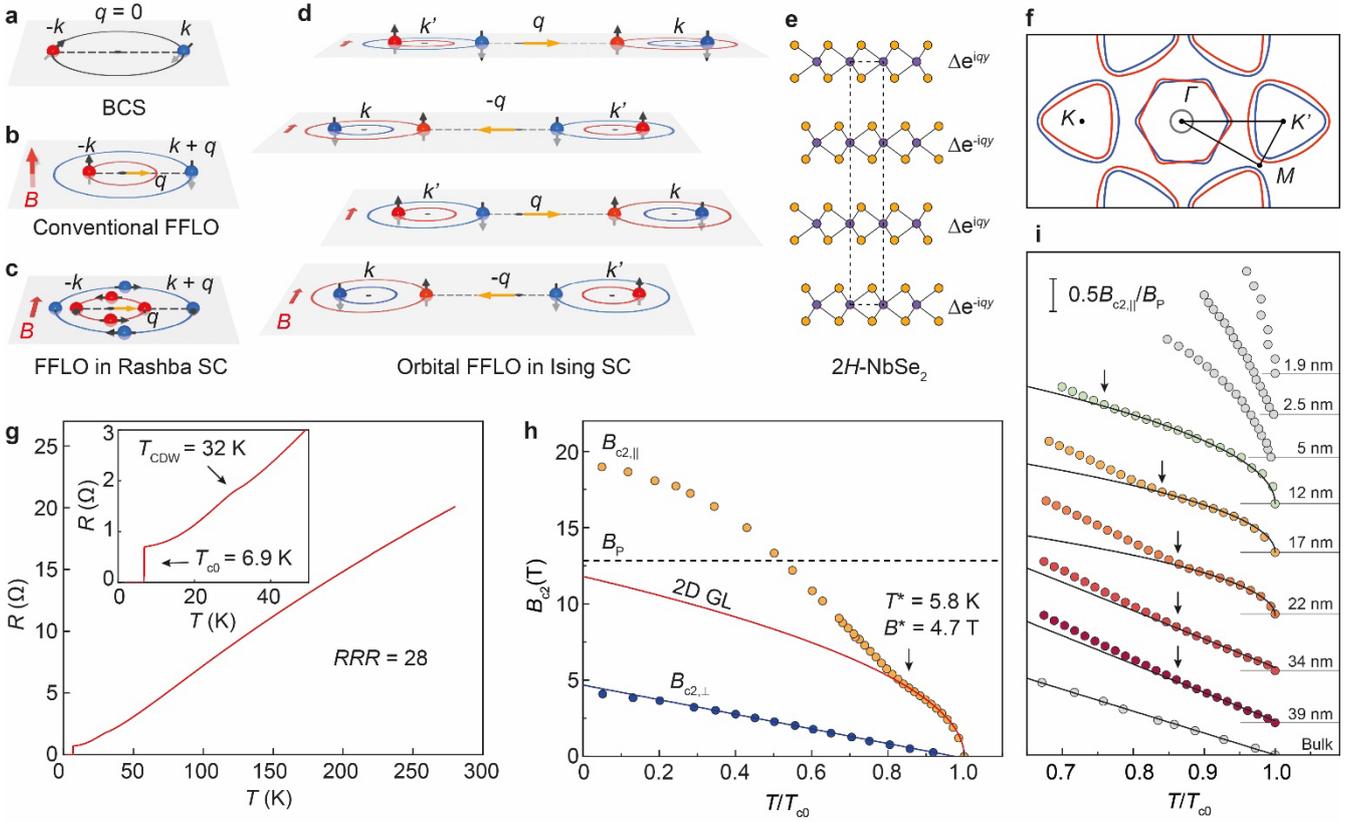

**Fig. 1: Superconductivity and possible pairing states in NbSe$_2$ multilayers. a,** BCS pairing with zero momentum. **b,** Conventional FFLO pairing with finite momentum $q$. The Zeeman effect of the external $B$ field induces spin imbalance. **c,** A representative of finite-momentum pairings in Rashba superconductors[4,16]. The Rashba SOC locks spin in the in-plane direction. The finite-momentum pairing is then induced by coupling between the Rashba SOC and the Zeeman effect of parallel fields. The sphere with an arrow represents electron and spin in panels b and c. **d,** The finite momentum pairing in 2$H$-stacking multilayer Ising superconductor. Strong Ising SOC locks spin to the out-of-plane directions and suppresses the Zeeman effect of an in-plane field. The orbital effect from the parallel $B$ field shifts the center of Fermi pockets away from the $K/K'$ point of the Brillouin zone. Pairing the $k$ and $k'$ electrons in shifted Fermi pockets of the $l$-th layer satisfies $k - k' = (-)^l q$. **e,** Spatial modulation of the superconducting order parameter in the orbital FFLO state of 2$H$-NbSe$_2$. **f,** Illustration of the Fermi surface of a NbSe$_2$ monolayer. Red and blue colors denote two spin states polarized up and down. The superconducting gap opens at the Nb-derived $K/K'$ (trigonal) and the $\Gamma$ (hexagonal) pockets. **g,** Temperature dependence of the resistance of a 17 nm thick flake showing large $RRR = 28$. The enlarged low-temperature part in the inset shows the CDW transition at $T_{CDW} = 32$ K and superconducting transition at $T_c = 6.9$ K. **h,**

Temperature dependences of the upper critical field for fields applied parallel (red, $B_{c2,\parallel}$) and perpendicular (blue, $B_{c2,\perp}$) to the 2D crystal plane. An upturn is observed at ($T^*$, $B^*$). The blue line is a fit using the 3D GL model for the upper critical field. The red line is a fit using the 2D GL model at $T > T^*$. **i,** The thickness dependencies on $B_{c2,\perp}/B_P$ as a function of $T/T_{c0}$. The abrupt upturn is observed in the intermediate thickness range between 10 ~ 40 nm. The error bars are smaller than the symbol size.

# Figure 2

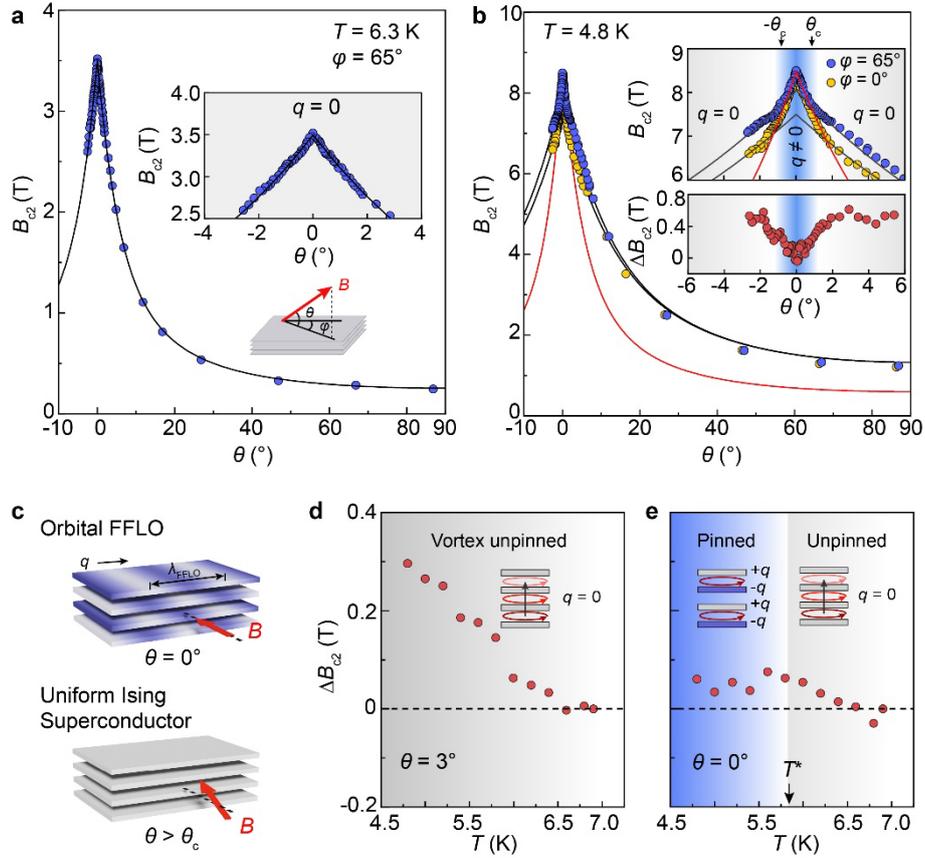

**Fig. 2: Upper critical fields and vortex dynamics in the orbital FFLO state.** The $I$ in 17 nm flake is kept along the direction as measurements shown in Fig. 3c-i. The $B_{c2}$ as a function of polar angle $\theta$ measured at **a,** $T = 6.3$ K $> T^*$ and **b,** $T = 4.8$ K $< T^*$. Insets are enlarged views near $\theta = 0°$. At $T = 6.3$ K, $B_{c2}$ shows one cusp shape that can be fitted by a single 2D Tinkham component for the whole $\theta$ range. At $T = 4.8$ K, the $B_{c2}$ fit requires an extra 2D-Tinkham component with a larger $B_{c2}$ for $|\theta| < \theta_c = 1°$. The enhancement of $B_{c2}$ at $|\theta| < \theta_c$ indicates the formation of the orbital FFLO phase. The lower insert of panel b shows $\Delta B_{c2} = B_{c2}(\varphi = 65°) - B_{c2}(\varphi = 0°)$ as a function of $\theta$. As shown later in Fig. 3c-i, $\varphi = 65°$ and $0°$ are two representative angles where the Lorentz forces are weak and strong, respectively. In the uniform phase at $|\theta| > \theta_c$, compared to the weak Lorentz force at $\varphi = 65°$, the stronger Lorentz force at $\varphi = 0°$ suppresses the $B_{c2}$ by activating the interlayer vortex motion, therefore causing $\Delta B_{c2} > 0$. The influence of the Lorentz force on $B_{c2}$ disappears at $|\theta| < \theta_c$, where the vortex motion is pinned after forming the FFLO state. **c,** Schematic for destroying the orbital FFLO when tilting the $B$ field to $|\theta| > \theta_c$. The stripes in the upper panel illustrate the real-space phase variation in the order parameters, showing wavelength $\lambda_{FFLO} = 2\pi/q$. Here, $q$ in the $l$-th layer is $\boldsymbol{q}_l \propto (-)^l(\boldsymbol{B} \times \boldsymbol{z})$, which is perpendicular to the $\boldsymbol{B}$ field. The influence of the Lorentz force, indicated by $\Delta B_{c2}$ is plotted as a function of temperature, showing the temperature dependence of the vortex pinning effect for the uniform phase at **d,** $\theta = 3°$ and the orbital FFLO phase at **e,** $\theta = 0°$, respectively. For $\theta = 3°$, the $\Delta B_{c2}$ gradually grows with decreasing temperature below $T^*$ due to the unpinned vortex motion in the uniform phase. However, for $\theta = 0°$, $\Delta B_{c2} \sim 0$ at $T < T^*$, indicating vortex pinning in the orbital FFLO phase.

# Figure 3

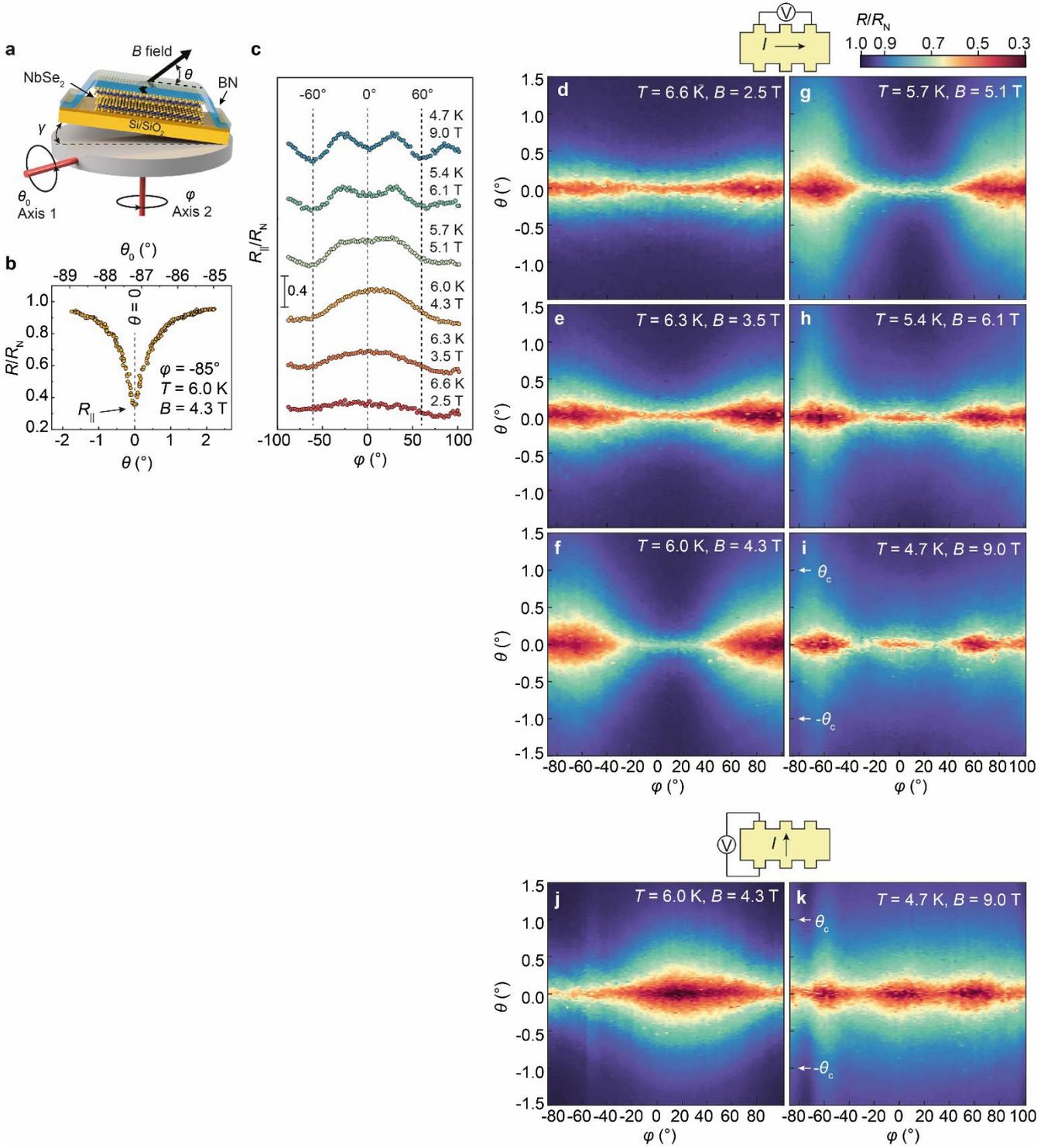

**Fig. 3: Six-fold anisotropy in the orbital FFLO state (measured in the 17 nm flake). a,** Schematic of a two-axis rotation stage with the device mounted in a small canting angle $\gamma$ that varies in each device installation. **b,** The polar angle $\theta$ dependence of magnetoresistance determines the $\theta = 0°$ orientation, where the measured $R$ is defined as $R_\parallel$. **c,** Magnetoresistance $R_\parallel(\varphi)$ in parallel $B$ fields. **d-i,** Mapping of $R(\varphi, \theta)$ after subtraction of a canting angle $\gamma = 0.78°$, showing a two-fold to six-fold anisotropy transition

across the tricritical point ($T^*$, $B^*$). The six-fold anisotropy can only be observed at $|\theta| < \theta_c$, which is consistent with the critical angle observed in Fig. 2b. **j-k,** Mapping of $R(\varphi, \theta)$ following rotation of $I$ to the orthogonal direction. The two-fold anisotropy shifts along with the direction of $I$, indicating the extrinsic origin of the two-fold anisotropy. In contrast, the six-fold anisotropy is independent of the $I$ directions, which is consistent with the intrinsic anisotropy of the orbital FFLO phase.

# Figure 4

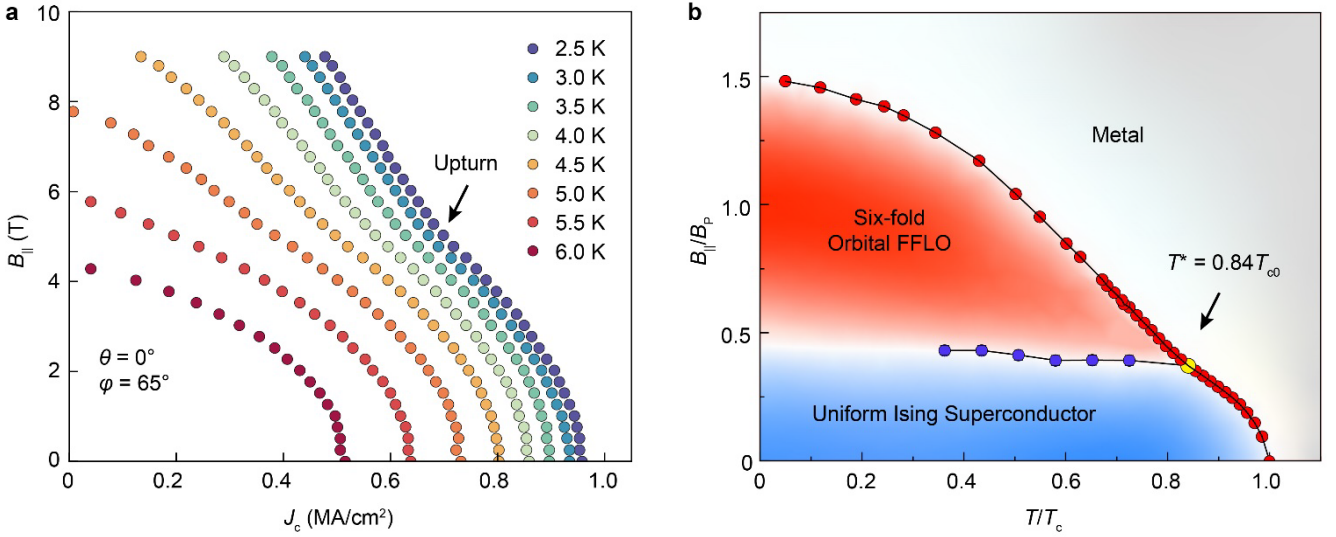

**Fig. 4: Phase diagram of the orbital FFLO state under parallel magnetic fields. a,** Magnetic field dependence of the critical current density $J_c$ at different temperatures. The upturns observed at low temperatures become broadened with increasing temperature. At $T \geq 6$ K, the upturn vanishes. **b,** Phase diagram of the orbital FFLO state. The metal-superconductor phase boundary (red and yellow dots) is determined by measuring the temperature dependence of $B_{c2,\parallel}$ as shown in Fig. 1h. The superconducting phase is divided into the uniform Ising superconductivity and the orbital FFLO state. The kinks observed in the $J_c(B)$ dependences (panel a) determine the boundary of the first-order phase transition from the uniform Ising superconductivity to the orbital FFLO states (blue dots). The Ising superconductivity is isotropic with a two-fold symmetry caused by the Lorentz-force-induced vortex motion. The orbital FFLO phase exhibits an intrinsic six-fold anisotropy. The error bars are smaller than the symbol size.

# Mothods

## Device fabrication and measurement

We mechanically exfoliate the NbSe$_2$ thin flakes from a bulk single crystal. The devices were made on a silicon substrate with an oxide layer of ~285 nm thick. The *h*-BN flakes were first exfoliated onto a PDMS substrate and dropped onto the NbSe$_2$ flake as the capping layer at room temperature. Both exfoliation and transfer processes were performed in an inert Ar atmosphere to prevent possible sample degradation. The electrodes (50 nm Au on 1 nm Ti) were deposited by electron beam evaporation after etching through the *h*-BN capping layer in CF$_4$ plasma. The thicknesses of the NbSe$_2$ flakes were determined by atomic force microscopy. The devices were mounted on a two-axis rotating stage (atto3DR). The angles were determined by potential meters, which have an accuracy better than 0.1°. Three SR830 DSP lock-in amplifiers measured the electrical transport. Two Keithley meters, a 2450 source/measurement unit and a 182 voltmeter, were used for DC $I-V$ measurement. If not specified, the data presented in this work were measured on a 17 nm thick flake with a constant current density of 0.003 MA/cm$^2$.

## Nullifying the canting angles due to the device installations

For a 2D superconductor, the angular dependence of the upper critical field $B_{c2}(\theta)$ is described by the 2D Tinkham formula[40]:

$$\frac{B_{c2}(\theta)\sin\theta}{B_{c2,\perp}} + \left(\frac{B_{c2}(\theta)\cos\theta}{B_{c2,\parallel}}\right)^2 = 1. \tag{1}$$

Here $\theta$ is the angle between the *B* field and the 2D plane. The $B_{c2,\perp}$ and $B_{c2,\parallel}$ are the upper critical fields when the field is perpendicular and parallel to the 2D plane, respectively.

For superconductors with strong anisotropy, *e.g.*, in our orbital FFLO system, even a tiny canting from the precise in-plane alignment can significantly decrease the $B_{c2}$ because of the large $B_{c2,\parallel}/B_{c2,\perp}$ ratio. It is worth emphasizing that even if we can tune the instrumental alignment close to being perfect, each device installation onto the measurement system can inevitably cause misalignment. In reality, there always exists a canting angle $\gamma$, contributed by both instrumental and installation misalignment.

In our angular measurements, this unavoidable $\gamma$ is eliminated by two-axis rotational mapping. We mount a 2D superconductor (having the yellow 2D plane) on a rotating stage (grey plane) with rotation axis 1 and 2. During the sample installation, a canting angle $\gamma$ is introduced between the sample and the rotation plane. The $\gamma$ can be canceled out by rotating axis 1 to align the sample plane with the $B$ fields. Nevertheless, once the 2D sample is parallel to the field ($\theta = 0$), rotating axis 2 will again induce an angle between the sample plane and $B$ field ($\theta \neq 0$), as shown in Extended Data Fig. 2, which varies as

$$\theta(\varphi, \theta_0) = \arctan[\tan\gamma \cdot \cos(\varphi + \Phi_2)] - (\theta_0 + \Phi_1), \qquad (2)$$

here $\Phi_1$ and $\Phi_2$ are the phases of the readings from axis 1 and 2. The $\gamma$, $\Phi_1$ and $\Phi_1$ become constants after the device installation. During the measurement, to keep the $B$ field parallel to the crystal plane at different $\varphi$, one must continuously tune the $\theta_0$. As a result, the canting angle $\gamma$ can be subtracted from the mapping of R($\varphi$, $\theta_0$) as shown in Extended Data Fig. 3. The misalignment of the $B$ field can shift the tricritical point $T^*$ to higher fields above the Pauli limit, which might have hindered the identification of orbital FFLO[41].

## Ginzburg-Landau model for the six-fold orbital FFLO

To understand the origin of the six-fold anisotropy under the in-plane $B$ field, we propose the following Ginzburg-Landau free energy of a multilayer $NbSe_2$ under in-plane magnetic field $B$ based on symmetries:

$$F = \sum_l \int d^2\mathbf{r} \left\{ \psi_l^* \alpha[(-)^l \mathbf{q}, \mathbf{B}] \psi_l - J(\psi_l^* \psi_{l+1} + c.c.) + \frac{1}{2}\beta |\psi_l|^4 \right\}, \qquad (3)$$

where $\psi_l(\mathbf{r}) = \int \frac{d^2\mathbf{q}}{(2\pi)^2} \psi_{l\mathbf{q}} e^{i\mathbf{q}\cdot\mathbf{r}}$ is the superconducting order parameter of layer $l = 1,2,\ldots$ in real space, $\mathbf{q} = -i\nabla$ is the in-plane momentum operator, $J > 0$ is the Josephson coupling, and $\beta > 0$ is the quartic coefficient of the order parameter.

The second-order coefficient $\alpha$ is a function of momentum $\mathbf{q}$ and magnetic field $\mathbf{B}$, and the layer-dependent alternating sign $(-)^l$ is due to the global inversion symmetry

$$I: \mathbf{q} \to -\mathbf{q}, \mathbf{B} \to \mathbf{B}, \psi_l \to \psi_{N+1-l}$$

where $N$ is the total number of layers. Together with the time-reversal symmetry, in-plane mirror symmetry, and in-plane rotation symmetry, we can determine the general form of $\alpha$ as

$$\alpha(\mathbf{q}, \mathbf{B}) = a(T - T_0) + b\mathbf{q} \cdot (\mathbf{B} \times \mathbf{z}) + cq^2 + \lambda_1 \mathrm{Re}(q_+^2 B_+^4) + \lambda_2 \mathrm{Re}(q_+^4 B_+^2) + \cdots, \qquad (4)$$

where $a, c > 0$ for the stability, and $q_+ = q_x + iq_y$, $B_+ = B_x + iB_y$. The parameters $a, b, c$, and $\lambda_{1,2}$ are determined by microscopic details of the system, such as density of states, Fermi velocity, SOC, Fermi surface anisotropy, and interlayer coupling. The dimensionless quantity $Jc/(bB)^2$ controls the phase transition in Eq. (4).

It is demanding to solve the complete problem with $N$ layers analytically. Therefore, without losing the essential physics, we examine the most straightforward multilayer system containing $N = 2$, forming the Bloch configuration. For this bilayer case, equation 4 can be solved perturbatively by treating $\lambda_{1,2}$ as the perturbations. For general $N$ layers, the order parameter can be obtained from the bilayer states as discussed later.

When $\lambda_{1,2} = 0$, by minimizing the free energy, we can find two kinds of superconducting phases near the superconducting-normal phase transition. The different Cooper pair momentum $q$ characterizes these two phases, hence different upper critical fields, or equivalently the $B$ field dependence of critical temperature $T_c$.

For $B < B^*$,

$$q = 0, \ T_c = T_0 + \frac{b^2}{4ac}(2|B^*|^2 - B^2).$$

And for $B \geq B^*$,

$$q_l = (-)^l \frac{b}{2c}(B \times z)\sqrt{1 - \left(\frac{B^*}{B}\right)^4}, \qquad T_c = T_0 + \frac{b^2}{4ac}\frac{|B^*|^4}{B^2}$$

And the tri-critical point $(T^*, B^*)$ is

$$B^* = \frac{\sqrt{2Jc}}{b}, \ T^* = T_{c0} - \frac{b}{2a}J,$$

The location of the tri-critical point $(T^*, B^*)$ is mainly determined by the interlayer Josephson coupling $J$. When the interlayer Josephson coupling is turned off, i.e., let $J = 0$, we find $T^* = T_{c0}$. This means that, at any finite field, each layer has Cooper pairs with alternating finite momentum $(-)^l q$, which is consistent with our previous argument. Also, we can determine the numerical constant relating $B^*$ and $\sqrt{Jc}/b$, consistent with our dimensional analysis discussed in the main text.

Next, we turn on the weak anisotropy by letting $\lambda_{1,2} \neq 0$, which are the perturbations. The anisotropic terms are all $q$-dependent and vanish at zero momentum, hence we find that the BCS phase at $q = 0$ is always isotropic. In contrast, the orbital FFLO phase with finite Cooper pair momentum shows six-fold anisotropy under an in-plane $B$ field.

For $B < B^*$,

$$q = 0, \ T_c = T_0 + \frac{b^2}{4ac}(2|B^*|^2 - B^2).$$

Whereas, for $B > B^*$,

$$q_l = (-)^l \frac{b}{2c}(\boldsymbol{B} \times \boldsymbol{z})\sqrt{1 - \left(\frac{B^*}{B}\right)^4}, \quad T_c = T_0 + \frac{b^2}{4ac}\frac{|B^*|^4}{B^2} + \lambda(B)\cos(6\varphi),$$

where $\lambda(B)$ only depends on the magnitude of the field, and $\varphi = \mathrm{atan}\,(B_y/B_x)$ is the polar angle denoting the direction of the in-plane $B$ field.

For the general case of an N-layer system, we expect the order parameter to form a Bloch-like configuration based on the bilayer states above, having the two-component order parameter

$$\Psi_l = \begin{pmatrix} \psi_{2l-1} \\ \psi_{2l} \end{pmatrix}.$$

And we then obtain the free energy of multilayers

$$F = \sum_l \int d^2\boldsymbol{r}\left\{\Psi_l^\dagger A \Psi_l - J\left(\Psi_l^\dagger \Psi_{l+1} + h.c.\right) + \frac{1}{2}\beta|\Psi_l^\dagger \Psi_l|^2\right\}, \tag{8}$$

$$A = \begin{pmatrix} \alpha_- & -J \\ -J & \alpha_+ \end{pmatrix}, \quad \alpha_\pm = a(T - T_0) \pm b\boldsymbol{q}\cdot(\boldsymbol{B}\times\boldsymbol{z}) + cq^2 + \cdots$$

Since the free energy above has the translation symmetry $\Psi_l \to \Psi_{l+1}$, according to the Bloch theorem, we expect our solution to satisfies $\Psi_{l+1} = e^{i\theta}\Psi_l$, or equivalently

$$\Psi_l = e^{il\theta}\Psi_0.$$

Here $\theta \in (-\pi, \pi]$ is the Bloch phase modulation parameter to be determined soon. With this ansatz of the order parameter, the free energy above becomes

$$F = \sum_l \int d^2\boldsymbol{r}\left\{\Psi_0^\dagger A \Psi_0 - 2J\Psi_0^\dagger \Psi_0 \cos\theta + \frac{1}{2}\beta|\Psi_0^\dagger \Psi_0|^2\right\}. \tag{9}$$

As the Josephson coupling is positive $J > 0$, when minimizing the free energy, we obtain $\theta = 0$. In other words, $\Psi_l = \Psi_0$ and every bilayer shares the same order parameter configuration. Therefore, the superconducting phases worked out from the bilayer NbSe$_2$ model can be extended to the multilayers.

## Excluding other possible scenarios for the upturn in $B_{c2}$ ($T$)

The upturn in $B_{c2}(T)$ is associated with out-of-plane translational and in-plane rotational symmetries breaking, which is consistent with the orbital FFLO model. Several other scenarios can also support the upturn in $B_{c2}(T)$, such as intrinsic two-gap superconductivity, the Takahashi-Tachiki effect, spin-triplet pairing, and 3D/2D dimensional crossover. In this section, the scenarios above are discussed and eventually ruled out.

**Two-gap superconductivity**

The superconducting gaps of $NbSe_2$ open at both $K$ and $\Gamma$ points for both hole- and electron-like pockets as shown in the ARPES results[42,43]. The superconductivity of the Se-derived (grey circle) $\Gamma$ pocket has a tiny gap[44] or is gapless[42,43]. The multiple gaps of different gap sizes may lead to kinks in both $B_{c2}(T)$[45] and $J_c(T)$[46]. On the other hand, in dirty two-gap superconductors, the upturn in $B_{c2}(T)$ is also expected due to inter-band scattering[45].

First of all, we can rule out the possibility of having the dirty two-gap scenario since our device shows superconductivity in the clean regime. Furthermore, the kink observed in $|\theta| < 1°$, as shown in the $B_{c2}(\theta)$ in Fig. 2b of the main text, cannot be described by the dirty two-gap model[45]. Second, the two-gap scenario would expect the temperature dependence of $J_c$ to show a kink when the second gap opens at $T < T_{c0}$ under the zero $B$ field[46]. As shown in Extended Data Fig. 7, the $J_c(T)$ measured at $B = 0$ T has a smooth dependence without showing kink as a function of temperature. Instead, the kinks in $J_c$ appear when the $B$ field is applied crossing the uniform superconductor/orbital FFLO phase boundary as shown in Fig. 4a. The phase boundary thus determined by $J_c(B)$ agrees well with the tricritical point that is determined by $B_{c2}(T)$ measurements. Hence, we can also safely rule out the two-gap scenario as the cause of the kinks observed in our devices.

**Takahashi-Tachiki effect**

The Takahashi-Tachiki (TT) effect describes a superconducting superlattice consisting of stacked superconducting bilayers. Two types of superconductors in the bilayers, the N and S layers, have the same $T_{c0}$, but different diffusion constants[47]. Both N and S superconductors are in the dirty limit. Therefore, the large/small diffusion constants (denoted as $D_N$ and $D_S$) lead to low/high $B_{c2}$ values. A new superconducting phase appears, when the $D_N/D_S$ exceeds a critical ratio, as an upturn in the temperature dependence of $B_{c2}$. The $B_{c2}$ upturn is associated with preferential nucleation of superconductivity in one of the two layers. Order parameters beyond the upturn are concentrated in the S layers with higher $B_{c2}$, hence, are more robust at high fields. As the physical picture of the TT effect is fundamentally different from our orbital FFLO state, they apply to very different regimes.

Firstly, the TT effect is developed explicitly for dirty superconductors. When both types of superconductors are in the dirty limit, where the competition and selective nucleation depend on the $D_S$ and $D_N$. In contrast, our sample is in the clean regime (SI 1), which is consistent with the high $RRR$ value (Fig. 1g). Furthermore, our system's large Ising spin-orbit coupling pins the spin in the out-of-plane direction, protecting the electron spins from being scattered by impurities. Therefore, the dirty regime assumption used in the TT model, where scattering determines the diffusion length[47], may not apply to our system.

Following the TT model, $B_{c2,S}/B_{c2,N}$ = 15 merely brings the upturn to $T^* \sim 0.5T_{c0}$[47]. To achieve $T^*$ = $0.84T_{c0}$ shown in Fig. 4, an unrealistically large $B_{c2,N}/B_{c2,S}$ is required. In NbSe$_2$, the highest in-plane $B_{c2}$ reported so far is ~$7B_P$, while the lowest $B_{c2}$ is found in pristine bulk for ~$1B_P$. Therefore, the largest $B_{c2}$ ratio accessible in NbSe$_2$ systems only reaches 7, which is less than half of $B_{c2,S}/B_{c2,N}$ = 15. As the ratio $B_{c2,S}/B_{c2,N}$ increases, increasing sharper upturns are expected. In contrast, experimentally, we observed a smooth upturn in $B_{c2}(T)$. In short, the relatively soft upturn and high $T^*$ observed in our uniform single crystal sample cannot be reconciled with the TT theory. It is worth noting that, the TT model has one more characteristic temperature close to $T_{c0}$, corresponding to a 2D/3D crossover[47]. In contrast, our 17 nm flake (Fig. 1h) shows no linear $B_{c2}(T)$ dependence, which is inconsistent with the TT theory.

The first-order transition described by the TT effect involves a competition between the coherence length $\xi$ and magnetic length $l_B = 1/\sqrt{2eB}$, which vary with temperature and $B$ field, respectively. Such competition leads to a first-order phase transition relying on both temperature and field, which appears diagonal in the phase diagram[47]. In our orbital FFLO picture, this first-order phase transition line is expected to have a weak temperature dependence. As shown in Eq. 4, the transition from uniform to orbital FFLO state is determined by $Jc/(bB)^2$ where a weak temperature dependence is expected from $J$. Indeed, experimentally, the first-order phase transition line depends weakly on temperature (Fig. 4b), which is inconsistent with the TT model.

**Spin-triplet pairing**

Due to the noncentrosymmetric crystal structure of monolayer transition metal dichalcogenides (TMDs), the spin splitting induced by Ising spin-orbit coupling may lead to mixed singlet and triplet pairing[48], which also enhances the $B_{c2}$[49]. Experimentally, the $B_{c2}$ of spin-triplet superconductivity characterizes a non-saturating temperature dependence due to the diverging paramagnetic limit. The non-saturating $B_{c2}$ has been reported in bulk spin-triplet superconductor UTe$_2$, violating the limits set by the orbital depairing effect and Pauli paramagnetism[50]. In TMDs, the spin-triplet superconductivity was not experimentally observed in monolayer MoS$_2$[10,51] and NbSe$_2$[28]. In our work, the $B_{c2}$ enhances moderately and saturates at low temperatures, which is inconsistent with the spin-triplet scenario. Furthermore, the spin-triplet pairing cannot explain the vortex pinning shown in Fig. 2. Therefore, we can rule out spin-triplet pairing as the candidate mechanism.

**Dimensional crossover from 3D to 2D**

An upturn in $B_{c2}$ has been widely observed in layered superconductors as a 3D to 2D crossover. Close to $T_c$, the $B_{c2}$ is described by the 3D Ginzburg-Landau model. Similar to the TT model, the phenomenological model valid in the scale of coherent length relies on the temperature dependence of coherence length. The fingerprint of such a crossover is a linear $B_{c2}$ dependence on temperature near the $T_{c0}$. The upturn due to dimensional crossover is well described by the KLB model[52]. For double-side gated bilayer $MoS_2$, a dimensional crossover related upturn is present due to the interlayer Josephson coupling[53].

In our data, the 3D/2D crossover scenario seems to capture the upturns of $B_{c2}$ in our 34 and 39 nm thick samples, which have linear $B_{c2}$ dependences on temperature (Fig. 1i). However, in 12 – 22 nm $NbSe_2$, the $B_{c2}$ shows a square root temperature dependence following the 2D Ginzburg-Landau model, suggesting a 2D superconductivity at high temperatures close to $T_{c0}$. As the 3D state is unavailable in the first place, the 3D/2D crossover cannot exist.

Furthermore, in our 17 nm flake, the 2D nature of superconductivity is confirmed in Fig. 2a and b, featuring cusp-shaped 2D angular dependence of $B_{c2}$ for the whole temperature range below $T_{c0}$. Therefore, the 3D/2D crossover cannot describe the observed $B_{c2}$ upturn.

## Excluding the nodal superconductivity and intrinsic gap anisotropy scenario for the six-fold anisotropy

When the magnetic field is applied along the $\Gamma$-$M$ lines, where the Ising SOC vanishes, the Zeeman effect can align the spins to in-plane direction, and thus closes the gap. The cooper pairing away from $\Gamma$-$M$ lines is still protected by Ising SOC and leads to a finite superconducting gap. Therefore, the nodal superconducting phase is suggested to exist above the Pauli limit[54], and exhibit six-fold anisotropy[55]. The recent work in a monolayer $NbSe_2$[56] reported a six-fold anisotropy in parallel magnetic fields at $T = 0.9T_{c0}$, which transforms into a two-fold anisotropy at $T = 0.5T_{c0}$. The six-fold anisotropy is suggested to be nodal superconductivity when the $B$ field is applied along the Γ-M lines, while the two-fold anisotropy is interpreted as nematic superconductivity.

Our results on multilayer $NbSe_2$ are very different compared with the theoretical[54] and experimental reports[56]. First of all, the six-fold superconducting phase appears well below the Pauli limit, which is not consistent with the theoretical prediction of nodal superconductivity[54]. Since the uniform/orbital FFLO phase boundary is determined by Josephson coupling, the boundary can be well below the Pauli limit[6].

On the other hand, the anisotropy transition in the monolayer NbSe$_2$[56] is essentially different from our results. In the monolayer, the anisotropy transition is driven by temperature. The six-fold nodal superconductivity is observed at $T = 0.9T_{c0}$, and the two-fold nematic phase is observed at $T = 0.5T_{c0}$. In contrast, the anisotropy transition in the orbital FFLO state is driven by magnetic fields. As indicated in Fig. 4b, the normal/orbital FFLO phase boundary has a weak temperature dependence. Furthermore, the two-fold anisotropy is extrinsically induced by Lorentz force. Therefore, the anisotropy transition is from isotropic to six-fold in our work, which is different from the report of nodal superconductivity[56]. With the above discussion, the scenario of nodal superconductivity can be ruled out in our work.

The intrinsic superconducting gap anisotropy of NbSe$_2$ has six-fold symmetry, as shown by ARPES and STM[42,43]. Nevertheless, our observation is a symmetry transition close to $T^*$ and $B^*$. Therefore, for $B < B^*$, where the intrinsic gap anisotropy is intact, we can only resolve the extrinsic two-fold symmetry due to the Lorentz force, excluding the intrinsic gap anisotropy as the origin of the six-fold symmetry found for $B > B^*$.

## Possible phase transition at lower temperatures

We noticed that pair density wave (PDW) superconductivity, which competes with the orbital FFLO phase, is an alternative candidate at lower temperatures and higher magnetic fields[7]. If it exists, one expects another upturn in $B_{c2}(T)$ at low temperatures where the transition from Ising FFLO to the PDW phase occurs. Meanwhile, following the theoretical prediction for conventional FFLO, a complex orbital FFLO state with multiple *q* may be energetically favored at low temperatures[57–59]. However, the present measurement shows that $B_{c2}(T)$ smoothly saturates at $T = 0.35$ K (Fig. 1h of main text) without any additional signature of PDW or multiple *q* phases. Our result, however, does not rule out the possibility that the PDW and multiple *q* orbital FFLO state may exist at an even lower temperature, where alternative configurations of broken translational symmetry may also be valid.

# Method references

065002 (2002).

59. Matsuda, Y. & Shimahara, H. Fulde–Ferrell–Larkin–Ovchinnikov State in Heavy Fermion Superconductors. *J. Phys. Soc. Japan* **76**, 051005 (2007).

60. Maloney, M. D., de la Cruz, F. & Cardona, M. Superconducting Parameters and Size Effects of Aluminum Films and Foils. *Phys. Rev. B* **5**, 3558–3572 (1972).

61. Kozuka, Y. *et al.* Two-dimensional normal-state quantum oscillations in a superconducting heterostructure. *Nature* **462**, 487–490 (2009).

62. Wang, B. Y. *et al.* Isotropic Pauli-limited superconductivity in the infinite-layer nickelate $Nd_{0.775}Sr_{0.225}NiO_2$. *Nat. Phys.* **17**, 473–477 (2021).

63. Takada, S. & Izuyama, T. Superconductivity in a Molecular Field. I. *Prog. Theor. Phys.* **41**, 635–663 (1969).
25

# Acknowledgements

This publication is part of the project TOPCORE (with project number OCENW.GROOT.2019.048) of the research programme Open Competition ENW Groot which is (partly) financed by the Dutch Research Council (NWO). P.W. acknowledges the research program "Materials for the Quantum Age" (QuMat) for financial support. This program (registration number 024.005.006) is part of the Gravitation program financed by the Dutch Ministry of Education, Culture and Science (OCW). O.Z. acknowledges the financial support of the CogniGron research center and the Ubbo Emmius Funds (University of Groningen). N.F.Q.Y. acknowledges the National Natural Science Foundation of China (Grant. No. 12174021) for the financial support. The high- field measurement was supported by HFML-RU/NWO-I, a member of the European Magnetic Field Laboratory (EMFL). It is part of the research programme of the Netherlands Organisation for Scientific Research (NWO) funded by the National Roadmap for Large-Scale Research Facilities.

# Extended figure 1

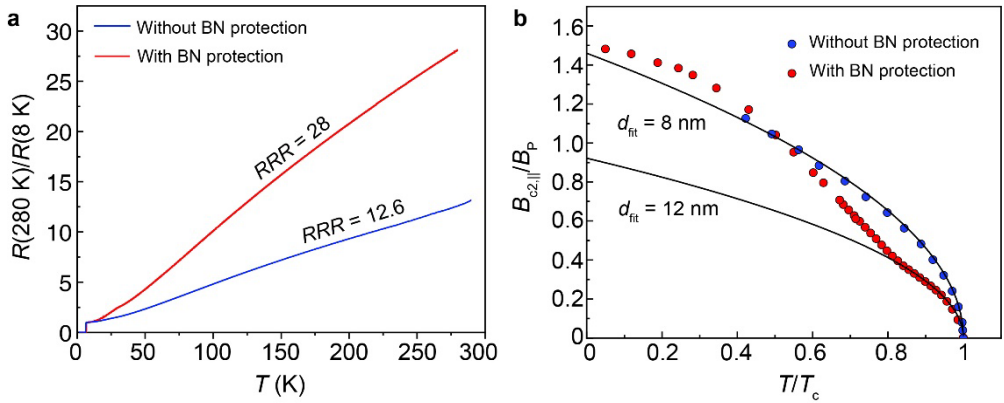

**Extended Data Figure 1 | Absence of the upturn in upper critical fields in a downgraded device.**

**a.** The temperature dependence of sheet resistances for two flakes showing $RRR$ = 12.6 and 28 for 11 and 17 nm thick flakes, respectively. **b.** The $B_{c2,\parallel}$ measured for the 11 and 17 nm thick flakes. The 2D Ginzburg-Landau (GL) fittings, *i.e.*, the solid black curves, yield thickness $d_{fit}$ = 8 and 12 nm, respectively. Overall, the thinner thicknesses obtained from the GL fittings are due to the protection from the Ising SOC, which becomes more robust in thinner flakes[60–62]. Close to the $T_{c0}$, the 11 nm flake shows a steeper temperature dependence of $B_{c2}$, consistent with its thinner thickness. Nevertheless, for the 17 nm flake with a larger $RRR$, an upturn in the $B_{c2}$ can be observed at $B = 0.36B_P$, indicating the orbital FFLO states, which eventually enhances $B_{c2}$ to exceed that measured in the 11 nm flake. As a larger $RRR$ indicates better sample quality, the contrasting behavior in the temperature dependence of $B_{c2}$ suggests that the absence of the orbital FFLO phase in the thin flake might be caused by the downgraded quality, which suppresses the finite-momentum pairing via scattering[13,63].

# Extended figure 2

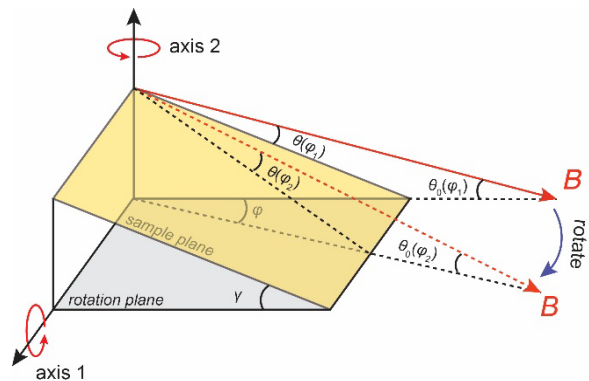

**Extended Data Figure 2 | Illustration of 2-axis rotation of a 2D sample in an external magnetic field with an installation canting angle γ.**

A 2D sample is mounted on a 2-axis rotational stage. The 2D surface of the sample (yellow plane) makes a canting angle $\gamma$ with respect to one of the rotation planes of the stage (grey plane). To simplify the discussion and isolate the effect of canting angle $\gamma$, we assume that the stage can make precise rotations so that, as shown in Main Text Fig. 3b, we can always align the sample plane precisely parallel to the external $B$ field. When this exact parallelism is aligned at a given $\varphi$, due to the canting angle $\gamma$, further rotation along the stage axis 1 or 2 can cause a correlation between $\theta$ and $\varphi$, which are labeled as different $\theta(\varphi)$ values.

# Extended figure 3

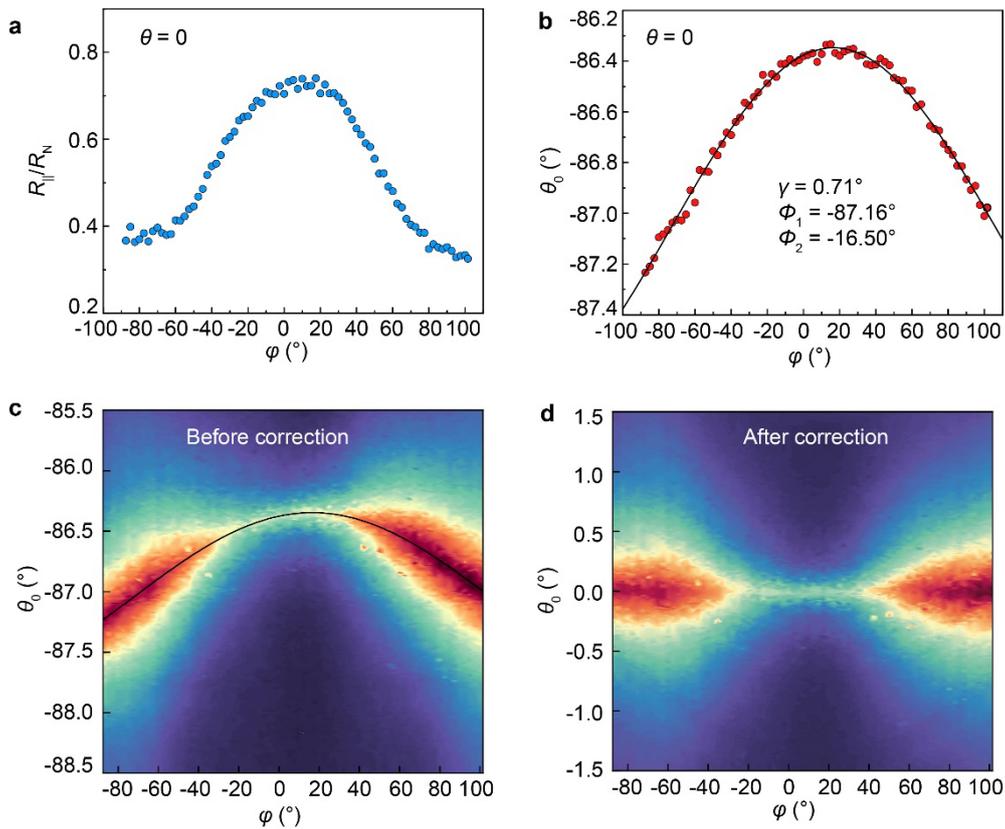

**Extended Data Figure 3 | Procedure for subtracting the canting angle γ.**

**a.** Magnetoresistance $R_\parallel(\varphi)$ in an in-plane $B$ field $B_\parallel$ (for the 17 nm device). **b.** Variation of $\theta_0$ (as defined in Extended Data Fig. 2) as a function of $\varphi$ when the $B$ field is adjusted to be parallel to the sample plane.

The solid black curve is fitting using Eq. 2 when $\theta = 0°$, which yields a canting angle $\gamma = 0.71°$. **c.** The data are shown in Main Text Fig. 2f before correcting the effect of canting angle $\gamma$. The black line is the same fitting that is shown in panel b. **d.** After correcting the canting $\gamma$, the magnetoresistance $R(\varphi, \theta)$ shows a two-fold anisotropy.

# Extended figure 4

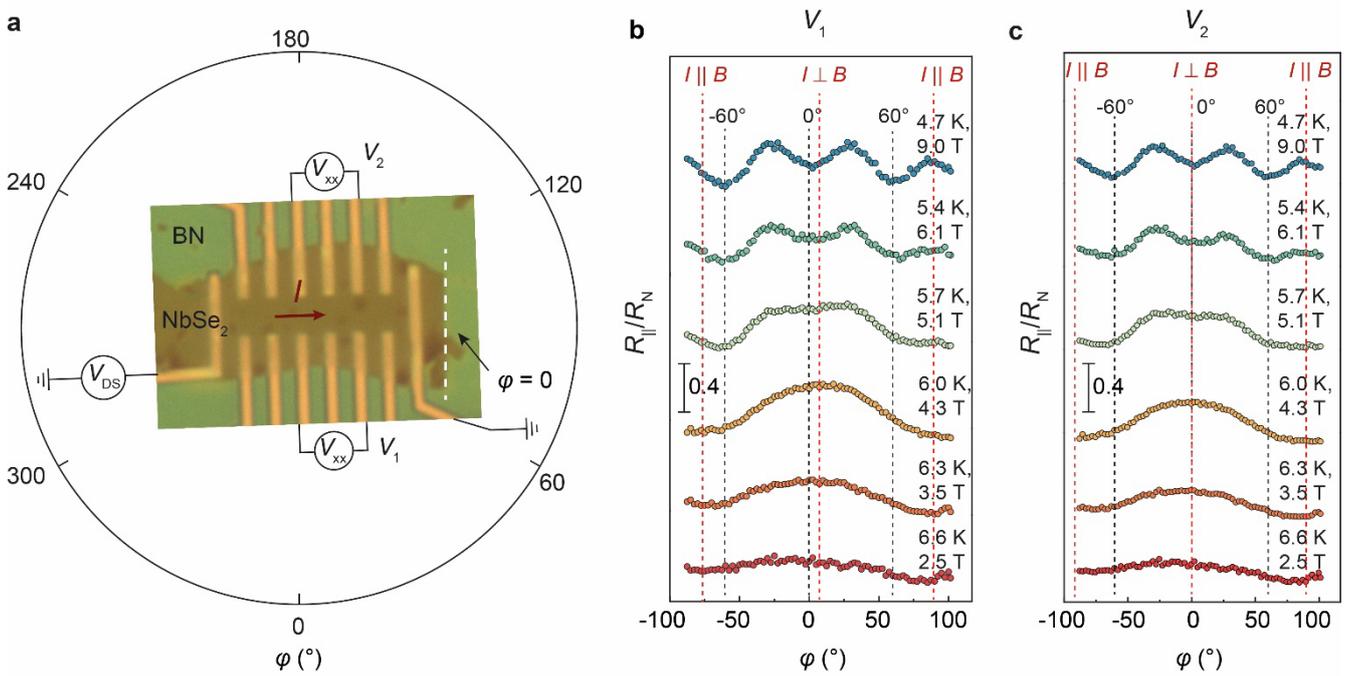

**Extended Data Figure 4 | Measurement configuration of the 17 nm device.**

**a.** Device orientation and the applied current direction. One crystalline direction of NbSe$_2$, as indicated by the white dashed line, is defined as $\varphi = 0$. **b, c.** Transport measurements using two sets of electrode pairs on two sides of the Hall bar show a small shift (~5°). It is consistent with the small deviation of $\varphi$ when changing the current direction in Fig. 3j of the main text.

# Extended figure 5

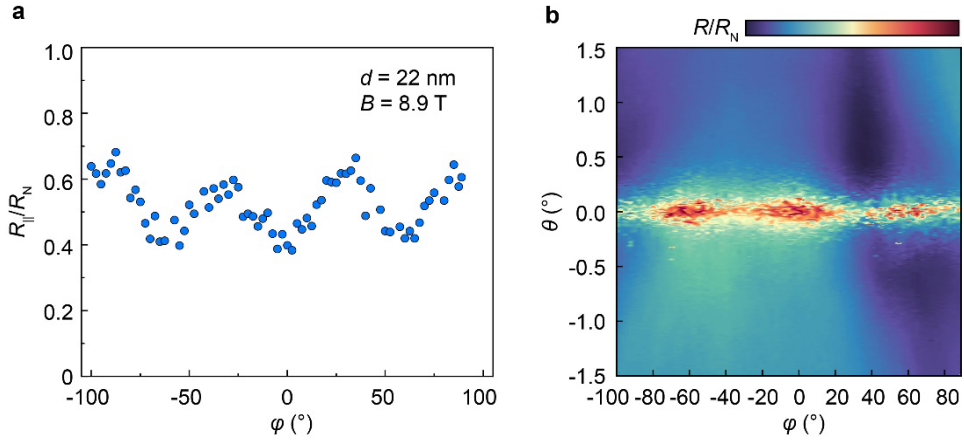

**Extended Data Figure 5 | Six-fold anisotropy of another multilayer NbSe₂ flake.**
The thickness of the flake is 22 nm. The anisotropy is measured at $B_\parallel$ = 8.9 T. **a.** The magnetoresistance $R_\parallel(\varphi)$ in the coexisting state shows a six-fold anisotropy in the $B_\parallel$ field. **b.** The mapping of $R(\theta, \varphi)$ exhibits the six-fold anisotropy when rotating the $\theta$ close to 0.

# Extended figure 6

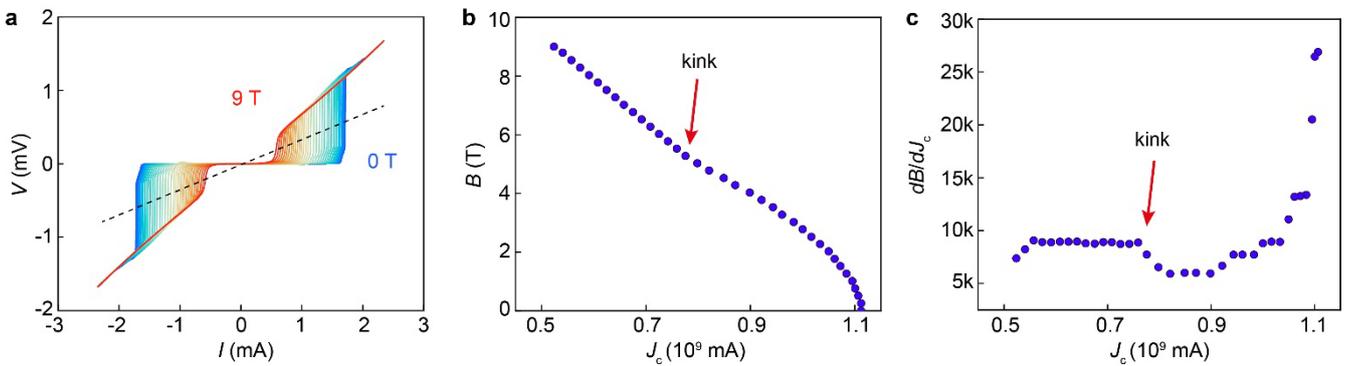

**Extended Data Figure 6 | Determination of the critical point where the upturn of $B_{c2}(J_c)$ occurs.**
**a.** An example of *I-V* measurements at different $B_\parallel$ fields at a fixed temperature. **b.** The critical current densities $J_c$ were extracted from panel **a**. The critical current density is determined as the point where *V/I* is half the normal resistance $R_N$ at $T$ = 10 K. **c.** The upturn in $B – J_c$ plot is determined by the kink in $dB/dJ_c$.

# Extended figure 7

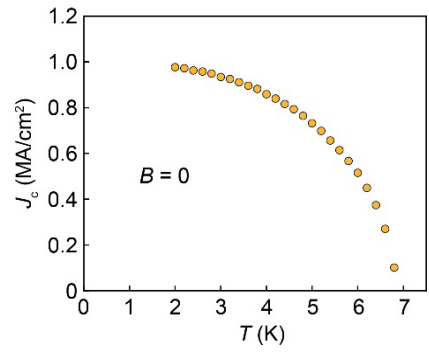

**Extended Data Figure 7 | Critical current density as a function of temperature under a zero magnetic field.**

At $B = 0$ T, no upturn was observed in the temperature dependence of $J_c$, ruling out the two-gap scenario as the cause of the upturns.